# Novel coupling between charge order and time-reversal-symmetry-breaking superconductivity

Quanxin Hu[1], Lingfeng Zhang[2], Yu Zheng[1], Yongwei Li[1], Qiheng Wang[1], Xinyu Liang[1], Baiqing Lv[1,3], Chi-Ming Yim[1], Takuto Kawakami[4], Vadim Grinenko[1,†], Xiao Hu[5,†], Hong Ding[1,6,7,†]

[1] *Tsung-Dao Lee Institute and School of Physics and Astronomy, Shanghai Jiao Tong University, Shanghai 201210, China*

[2] *Shanghai Key Laboratory of High Temperature Superconductors, Department of Physics, Shanghai University, Shanghai 200444, China*

[3] *Zhangjiang Institute for Advanced Study, Shanghai Jiao Tong University, Shanghai 200240, China*

[4] *Department of Physics, School of Science, The University of Osaka, Osaka 565-0871, Japan*

[5] *Institute for Quantum Science and Technology, and Department of Physics, Shanghai University, Shanghai 200444, China*

[6] *Hefei National Laboratory, Hefei 230088, China*

[7] *New Cornerstone Science Laboratory, Shanghai 201210, China*

[†]Corresponding author. Email: vadim.grinenko@sjtu.edu.cn; hu_xiao@shu.edu.cn; dingh@sjtu.edu.cn

**The interplay between charge-density waves (CDWs), which break translational symmetry, and spatially homogeneous superconductivity, which breaks global U(1) gauge symmetry, can give rise to an intriguing phenomenon: the pair-density wave, characterized by a spatial modulation of the superconducting order parameter. Yet how CDWs couple to unconventional superconducting states—particularly those with time-reversal symmetry breaking (TRSB)—remains largely unexplored. Here, using scanning tunneling microscopy on heavily hole-doped $Ba_{1-x}K_xFe_2As_2$, which hosts an $s \pm is$ superconducting state, we reveal a previously unobserved coupling between a surface CDW and TRSB superconductivity. Experimentally, the TRSB superconductivity imparts "chirality" to the CDW, which manifests as commensurate domains separated by domain walls with π-phase slips—forming what we term a *bipolar CDW*. The domain walls delineate TRSB domains of opposite chirality, consistent with spontaneous breaking of $U(1) \times Z_2$. Supported by theoretical modelling, we construct a framework in which a hidden interfacial pair-density modulation (PDM) mediates a linear coupling between the surface CDW and interband Josephson currents of TRSB superconductivity. Crucially, the theory shows that realizing this linear coupling requires a controlled global phase difference $\delta\varphi = \pi/2$ between the PDM and CDW states. Our results uncover a previously overlooked connection between charge ordering and TRSB superconductivity, opening a pathway to explore intertwined quantum orders in iron-based superconductors and other strongly correlated systems.**

Intertwined quantum orders are hallmark features of strongly correlated electron systems and often give rise to emergent phenomena beyond the framework of single-particle theories. Among them, the interplay between charge-density waves (CDWs) and superconductivity has attracted longstanding attention, particularly in high-temperature superconductors[1–9], heavy fermion superconductors[10], kagome superconductors[11,12] and transition-metal dichalcogenide[13]. CDW is a collective electronic state characterized by a periodic modulation of the charge density that breaks the translational symmetry of the crystal, whereas conventional superconductivity is a spatially uniform Cooper-pair condensate that breaks global $U(1)$ gauge symmetry. Despite significant efforts, most existing studies of the interplay between charge order and superconductivity have primarily focused on the simplest case of spatially homogeneous Cooper-pair condensates. In such scenarios, when Cooper-pair condensates develop in the presence of a periodic CDW potential $V(r)$, the finite center-of-mass momentum imposed by the CDW wave vector $\pm Q_{CDW}$ will lead to a spatial modulation in the Cooper-pair density[14]. This effect suggests a natural tendency toward pair density wave (PDW) formation, where the superconducting order parameter acquires spatial modulation with $Q_P = \pm Q_{CDW}$ (Fig. 1a). By contrast, a broader — and potentially richer — regime remains largely unexplored: multicomponent, time-reversal symmetry breaking (TRSB) superconductors (e.g. $s \pm is$), whose complex order parameters, with inter-component couplings, could couple to charge order in qualitatively distinct and phase-sensitive ways not available in a homogeneous condensate (Fig. 1b). To date, however, the scarcity of materials unambiguously exhibiting coexisting TRSB superconductivity and CDW order has precluded systematic investigations of their couplings.

Iron-based superconductors exhibit a rich landscape of intertwined electronic orders, including nematicity[15], spin[16], and charge orders[17–21], especially in an archetypal compound $Ba_{1-x}K_xFe_2As_2$ (BKFA). The crystal structure of BKFA is shown in Fig. 1d. Due to the polar nature of the cleaved surfaces, the near-surface carrier density depends sensitively on the termination layer[18,22]. Surface-sensitive spectroscopic studies have revealed a CDW but without superconductivity on the As-terminated surface of heavily hole-doped BKFA, attributed to the nesting between saddle points[18]. Concurrently, BKFA is a multiband superconductor whose Fermi surface evolves strongly with K concentration. Angle-resolved photoemission spectroscopy (ARPES) studies have revealed that with increasing K doping, BKFA undergoes a Lifshitz transition, during which the electron pocket at the X point is pushed above the Fermi level and transforms into a hole-like pocket[23]. The Lifshitz transition of the Fermi surface topology is accompanied by a change in the pairing symmetry from nodeless $s_\pm$ to nodal $s_\pm$[24]. Within a narrow doping dome of 0.70 < x < 0.85 ("magic doping" region), Muon spin rotation ($\mu$SR) and thermoelectric measurements have provided compelling evidence for a complex order parameter with the interband phase difference being neither 0 nor $\pi$ [25–29]. Such a complex pairing state, termed $s \pm is$, spontaneously breaks $U(1) \times Z_2$ symmetry. The two $Z_2$ degenerate ground states possess two different Josephson-current-like 'loops' in momentum space: either from $\phi_1 \to \phi_2 \to \phi_3 \to \phi_1$, or alternatively, $\phi_1 \to \phi_3 \to \phi_2 \to \phi_1$ the first component to the second, from the second to the third and from the third back to the first, or in the opposite order (Fig. 1c). Complex conjugation reverses the circulation, mapping the state to its distinct $Z_2$ partner; hence time-reversal symmetry is spontaneously broken.

Within the magic doping regime, the coexistence of a surface CDW on the As-terminated surface and a bulk TRSB superconducting state of the $s \pm is$ type naturally forms a connate heterostructure, in which the CDW and superconductivity are strongly coupled yet spatially separated. This configuration provides a unique platform to investigate the interplay between TRSB superconductivity and CDW order—an intriguing possibility that has remained largely unexplored, yet may hold the key to uncovering the microscopic origin of time-reversal symmetry breaking in iron-based superconductors. Here, using scanning tunneling microscopy and spectroscopy (STM/S), we report the direct observation of the as-yet-discovered interplay between CDW and a TRSB superconducting state in heavily hole doped BKFA. We identify an unusual type of CDW, characterized by a $\pi$-phase shift between neighboring domains coupled to TRSB superconducting states with opposite chirality. We designate this state as a *bipolar CDW*. The experimental findings, supported by theoretical modelling, indicate that the emergence of this unconventional CDW requires the presence of hidden interfacial pair-density modulations (PDMs) whose spatial modulation matches the surface-CDW wave vectors but whose global phase differs by $\delta\varphi = \pi/2$; these PDMs, in turn, mediate a linear coupling between the surface CDW and the interband Josephson currents associated with distinct superconducting components.

**Domain walls (DWs) on the As-terminated surface**

First, we focus on $Ba_{0.23}K_{0.77}Fe_2As_2$ (hereafter referred to as BKFA77), in which the superconducting state spontaneously breaks the $U(1) \times Z_2$ symmetry[26,27]. Due to the polar nature of the cleaved surfaces in BKFA, the cleaving process often induces surface reconstruction and partial loss of Ba/K atoms in order to maintain charge neutrality[30,31]. To minimize such surface artifacts and preserve the intrinsic electronic structure, we cleaved BKFA single crystals *in situ* at 14 K under ultrahigh vacuum before inserting them into the STM stage. By carefully manipulating the residual Ba/K adatoms with the STM tip, we are able to expose a pristine, atomically resolved As-terminated surface. Figure 1e shows the spatially resolved differential conductance map, d$I$/d$V$ ($r$, V), acquired at $V$ = -10 mV, together with its two-dimensional Fourier transform (Fig. 1f), clearly reveals a checkerboard-like CDW with a spatial periodicity of 2$a_{As}$ × 2$a_{As}$. At the same time, superconductivity is entirely suppressed on this As-terminated surface[18].

In addition to the long-range commensurate CDW order, we observe the presence of bright linear features with enhanced local density of states in the d$I$/d$V$ ($r$, V) map at $V$ = -10 meV (Fig. 1e). While the CDW modulation appears locally uniform across the surface, a closer inspection reveals that the CDW undergoes a π-phase shift across these bright lines, as indicated by the orange dashed markers in Fig. 1e. Across these lines, the CDW intensity inverts, switching from a maximum to a minimum (Fig. 2b). These features define domain walls separating CDW domains of opposite phase. This phase shift is further corroborated in reciprocal space. The Fourier transform of the differential conductance map (Fig. 1f) shows a clear splitting of the primary (2 × 2) CDW peaks, highlighted by blue squares, in sharp contrast to the As atomic lattice, which contributes only a single well-defined intensity peak marked by the red squares. Such splitting directly reflects the presence of anti-phase boundaries between adjacent domains[32] (see Extended Data Fig. 1 for details). Notably, the orientations of these domain walls exhibit randomly curved trajectories with respect to the underlying crystallographic axes, indicating that their formation is not pinned by the lattice symmetry. In reciprocal space, their signatures reside near $q_0 \approx 0$ (see Extended Data Fig. 2 for details). This behavior stands in stark contrast to transition-metal dichalcogenides such as 1T-TaS$_2$[33] and 1T-TiSe$_2$[34], where domain walls typically form straight along the CDW unit-vector direction.

The charge density field has amplitude $\rho_{q_i}^0(r)$ and phase $\varphi_i(r)$, both of which can vary as a function of position in real space: $\rho_{q_i}(r) = \rho_{q_i}^0(r) \cos\left(q_i^{CDW} \cdot r + \varphi_i(r)\right), i = 1, 2$. To expose the information encoded in these fields, we employ a two-dimensional lock-in technique to analysis the amplitude and phase of this periodic modulations[35] (see methods for more details). Extended Data Fig. 3 shows the amplitude $\rho_{q_i}^0(r)$, $\cos\left(q_i^{CDW} \cdot r + \varphi_i(r)\right)$, relative phase $\varphi_i(r)$ and phase gradient for $q_1^{CDW}$ and $q_2^{CDW}$. The relative phase $\varphi_i(r)$ maps are displayed in the interval $[-\pi, \pi]$ and reveal several topological defects of the CDW order, visible as $2\pi$ phase windings (white circles), as shown in Extended Data Fig. 3c and 3g. Phase gradients are evaluated on the unwrapped phase maps. The bright lines reveal the $\pi$ shift domain walls.

**Spectroscopic characterization and time dependence of DWs**

Next, we focus on the basic spectroscopic characterization of the CDW and domain wall regions. Figure 2c presents a high–spatial-resolution d$I$/d$V$ linecut measured along the red line marked in Fig. 2a, traversing a domain wall. The spatially averaged d$I$/d$V$ spectra taken from the commensurate CDW region (red curve) and the domain wall region (blue curve) are shown in Fig. 2d for direct comparison. These measurements reveal distinct spectroscopic signatures between the two regions. At the domain wall, the density of states (DOS) near the Fermi level is noticeably suppressed, while the spectral weight at negative bias shifts to higher intensity. This redistribution of spectral weight highlights a clear modification of the local electronic structure at the domain wall.

Beyond their unconventional morphology, the domain walls uncovered here display an entirely unexpected property: they are dynamic. As revealed in Fig. 2e, we recorded constant-current maps d$I$/d$V_{cc}$ (r, $V$ = –10 mV) from the same region every 10 h and discovered that the distribution of domain walls undergoes a slow temporal drift—clear evidence of dynamical behavior persisting even at the ultralow temperature of 0.3 K. Such behavior is unprecedented: in other correlated materials, the suppression of commensurate CDW order by heat, pressure, doping, or ultrafast optical and electrical pulses can give rise to quasi-metallic metastable phases separated by discommensuration domain walls[34,36–40]. Yet once established, these domain walls are invariably frozen at low temperatures, unless perturbed externally. The persistence of dynamics in our system therefore points to an extraordinarily delicate energy landscape, hosting two nearly degenerate ground states, as illustrated in Fig. 2f with two distinct colors. Moreover, the domain wall configuration is highly sensitive to both magnetic field (Extended Data Fig. 4) and temperature (Extended Data Fig. 5). Applying an out-of-plane field reorganizes the DW configuration. After cycling the field back to zero, the resulting distribution no longer matches the initial zero-field state, evidencing pronounced magnetic-history dependence. Likewise, thermal annealing to 27 K (above the bulk superconducting transition) followed by re-cooling yields a markedly different DW configuration, indicating strong thermal-history dependence. These remarkable features raise the intriguing possibility that their formation is intimately connected to the intrinsic degeneracy of the underlying TRSB superconducting states — namely, the $s + is$ and $s - is$ states.

**Doping dependence of DWs**

To investigate this correlation further, we performed STM measurements on a series of BKFA single crystals with varying potassium concentrations: x = 0.77, 0.79, 0.82, 0.86, 0.94, and 1.00 (Fig. 3a). The superconducting transition temperatures ($T_c$) of these samples, as used in STM experiments, were independently determined by A.C. susceptibility measurements (Fig. 3b; see Methods). Among these, the samples with x = 0.77, 0.79, and 0.82 lie within the magic doping region with TRSB superconducting state $s \pm is$, where the superconducting order is expected to break both $U(1) \times Z_2$ symmetries. In contrast, samples with x ≥ 0.86 fall outside the TRSB dome and are believed to host a conventional $U(1)$-breaking nodal $s_\pm$ pairing state without additional discrete symmetry breaking[27].

Figure 3c presents a series of constant-current differential conductance maps, d$I$/d$V_{cc}$ (r, $V$ = –10 mV), acquired on the As-terminated surfaces of these samples[18]. A commensurate 2 × 2 checkerboard-like CDW is observed consistently across all doping levels. However, DWs—identified as π-phase-shifted CDW regions—emerge only in the TRSB doping range (x = 0.77–0.82), and are entirely absent in samples with x = 0.86, 0.94, and 1.00 (see more measurements in Extended Data Fig. 6). These findings rule out the possibility that the DWs arise from the CDW itself—such as lattice distortions, defects, or conventional CDW phase slips[33,36,37,41]. Instead, the robust and reproducible emergence of π-phase-shifted DWs exclusively within the TRSB superconducting regime strongly indicates that they are rooted in the underlying $s \pm is$ pairing state.

**Origin of the DWs**

Having identified that the DWs are intimately connected to the superconducting pairing state in the bulk, a fundamental question arises: why does the TRSB $s \pm is$ pairing state induce DWs in the surface CDW, and why is a π-phase shift observed across these DWs? In multiband superconductors, the emergence of the $s \pm is$ state stems from the competition among distinct interband pairing channels[28,42]. Without loss of generality, by setting the phase of the first superconducting component $\phi_1 = 0$, the frustrated ground state exhibits interband phase differences $\phi_{12}$ and $\phi_{13}$ that are neither 0 nor $\pi$, as illustrated in the inset of Fig. 1c. This frustration leads not only to a complex phase landscape, but also to a discrete two-fold ground-state degeneracy. Specifically, the system supports two energetically degenerate but distinct states—$s + is$ and $s - is$—which are related by time-reversal transformation and spontaneously break $Z_2$ symmetry. According to the Kibble–Zurek (KZ) mechanism[43–45], topological defects (or DWs) associated with the spontaneous breaking of a discrete $Z_2$ symmetry are expected to form upon cooling through such a symmetry-breaking phase transition. Therefore, the most plausible origin of the DWs observed in the surface CDW is the presence of bulk $s \pm is$ pairing domains, whose phase discontinuities are projected onto the surface as π-phase-shifted CDW DWs.

Based on this scenario, we constructed a phenomenological Ginzburg–Landau-Wilson (GLW) model incorporating a surface CDW and a multicomponent superconducting order that breaks time-reversal symmetry. The model considers the simplest coupling between the CDW order parameter and an inter-component Josephson-current-like term (see methods for more details). For simplicity, an isotropic three-component superconductivity is presumed and $T_c^{SC} > T_c^{CDW}$. Minimizing the uniform GLW energy density, it is found that the TRSB superconductivity order parameter sets in a parabolic way of $\sqrt{1 - T/T_c}$, as displayed in Fig. 4a. In a stark contrast, the charge order is rapidly suppressed upon heating and develops in a linear way from the same critical temperature $T_c$, where $U(1) \times Z_2$ symmetries and the translational symmetry are broken simultaneously. Experimentally, the extracted CDW-peak intensity is in excellent agreement with this GLW phenomenology[18]. Extended Data Fig. 7 presents sequences of constant-current differential-conductance maps $dI/dV_{cc}$ (r, V) at V = -10 mV for $Ba_{0.23}K_{0.77}Fe_2As_2$ (BKFA77). The CDW contrast is rapidly suppressed and becomes indistinguishable by 13 K (blue circles), whereas the domain-wall (DW) signal—linked to TRSB superconductivity—persists up to a higher crossover temperature $T^*$ and vanishes only near 19.0 K (BKFA77), indicative of quartic correlations surviving above $T_c$[26,46]. We performed numerical calculations for the situation where two TRSB superconductivity states coexist in the system and are separated by an interface by imposing appropriate boundary conditions on superconductivity (white dashed lines in Fig. 4b), and check the stable wavefunction of the whole system which minimizes the free energy. As shown in Fig. 4b, accompanying the reversal between the two TRSB superconductivity states imposed by boundary conditions, the 2×2 charge order slides by a half-wave length across the interface, successfully capturing the main features observed in our experiments (see more details in Extended Data Fig. 8). The agreement between theory and experiment indicates that the linear coupling between the two orders plays a crucial role in this system.

**Discussion**

The Ginzburg–Landau–Wilson (GLW) framework is powerful for unveiling phenomena involving intertwined orders with distinct symmetries. In the present context, however, a linear coupling between the CDW order parameter and an inter-component Josephson-current–like term is highly unusual. Re-examining the connate heterostructure, we emphasize that the surface hosts only CDW while the bulk hosts only TRSB superconductivity—that is, the two orders are strongly coupled yet spatially separated. Inevitably, there exists an interfacial transition region where both order parameters are finite. In such a region, the finite center-of-mass momentum imposed by the CDW wave vector $\pm Q_{\text{CDW}}$ endows Cooper pairs with a spatial phase gradient, driving a modulation of the pair density—i.e., a pair-density modulation (PDM) that emerges naturally. In general, the (free-) energy density for such a coupling should take the form

$$(\phi \psi_k)^* \zeta_j + (\phi \psi_k) \zeta_j^*, \qquad 1)$$

where $\phi$ is the CDW order parameter for the charge order on the top surface, $\psi_k$ denotes the $k$-th uniform bulk superconductivity component, and $\zeta_j$ is an interfacial Cooper-pair density modulation arising from the $j$-th bulk superconductivity component. The pair density modulations and the 2 × 2 charge order share the same wave numbers $\boldsymbol{Q}_1 = (\pm q, 0)$ and $\boldsymbol{Q}_2 = (0, \pm q)$ with $q = \pi/a$. The position shift of $a/2$ between As atoms and the Fe atoms in the lateral directions of the lattice structure yields a phase difference $\delta \varphi = \pi/2$ between the surface CDW and the pair-density modulation (PDM), i.e. the two orders are locked in quadrature, as displayed in Figs. 4c and 4d (see more details in Extended Data Fig. 9). A comparable global phase offset between the PDM and CDW states has likewise been reported in other systems[47,48]. The coupling energy term in Eq. (1) then gives the linear coupling term

$$i\beta\phi(\psi_k^*\psi_j - \psi_k\psi_j^*) = 2\beta\phi|\psi_j||\psi_k|\sin(\theta_j - \theta_k) \qquad 2)$$

where a real CDW order parameter is taken as always possible and the real number $\beta$ reflects that the PDM is a residual of bulk superconductivity (see methods for more discussions).

Intriguingly, the charge order and the multi-component superconductivity couple to each other in the form of Josephson effect, which should manifest itself in TRSB multi-component superconductors[28]. The coupling linear in $\phi$ in Eq. (2) yields the possibility of a novel locking between the 2x2 charge order and the TRSB superconductivity, namely when the superconductivity state is reversed into its TRS partner with opposite inter-component Josephson currents, the charge order on the top As atom layer has to slide by half-wave length in order to keep the low (free-)energy, such that the sign change in Josephson currents is to be compensated by the sign change of $\phi$. As a result, the originally uniform surface CDW develops distinct "chiralities" depending on the underlying bulk superconducting domain. When crossing a domain wall between two $s \pm is$ states with opposite chiral Josephson-current-like loops, the surface CDW undergoes a $\pi$ phase shift. We refer to this unconventional CDW as a "bipolar CDW", emphasizing its correlation with the chirality of the bulk superconducting state.

In summary, we demonstrate a hitherto unknown coupling between charge order and unconventional superconductivity that breaks time reversal symmetry. A hidden interfacial pair-density modulation (PDM) mediates a linear coupling between the surface CDW and interband Josephson currents of the bulk TRSB superconductivity, through which the bulk TRSB order imprints a spatial modulation onto an otherwise uniform surface CDW, enforcing alternating chiralities and producing π-phase shifts across superconducting domain walls (Fig. 4e and Extended Data Fig. 10). We designate this unconventional CDW a bipolar CDW. This work not only unveils a novel manifestation of TRSB superconductivity in iron-based compounds, but also reveals a previously unobserved CDW–TRSB superconductivity coupling, thereby establishing a new paradigm for investigating intertwined quantum orders in iron-based superconductors and other strongly correlated systems.

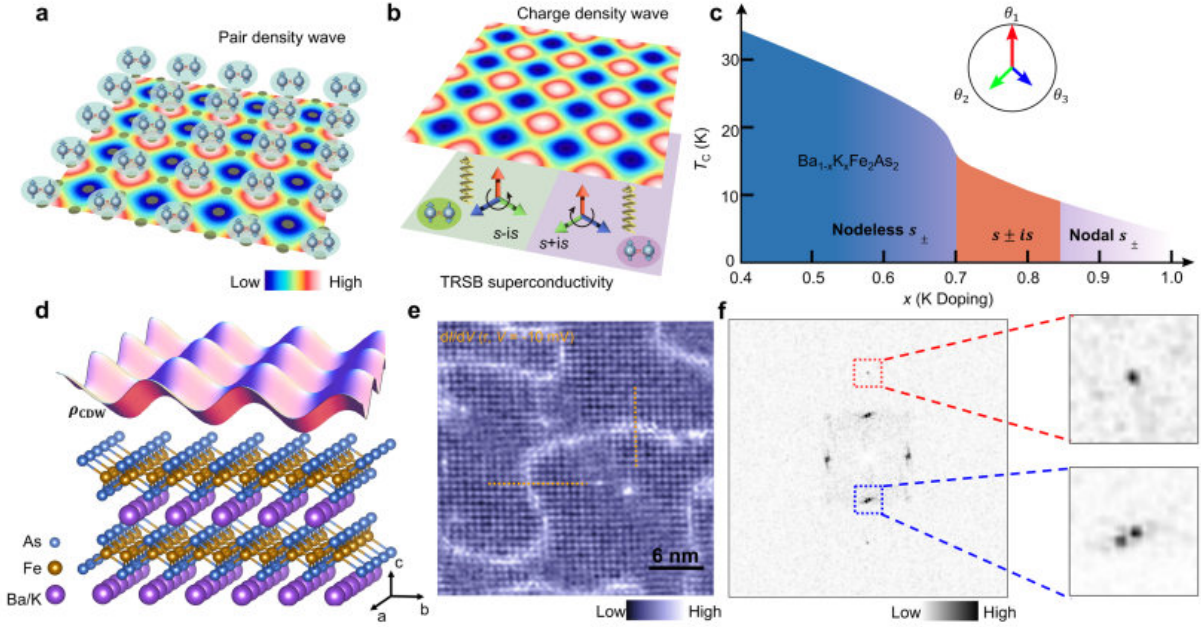

**Fig. 1 | Domain walls on the As-terminated surface. a,** When Cooper pairs condense in the presence of a CDW, the Cooper-pair density acquires a spatial modulation. This effect naturally favors the formation of a pair-density wave, where the superconducting order parameter acquires spatial modulation with $Q_P = \pm Q_{CDW}$. **b,** Schematic illustration of the coupling between multicomponent, time-reversal symmetry-breaking superconductivity and charge order. Such an interplay remains largely unexplored. **c,** Schematic phase diagram illustrating how the superconducting transition temperature ($T_c$) and pairing symmetry evolve with hole doping x. With increasing K concentration, the system undergoes a sequence of pairing states: from a nodeless $s_\pm$ state, to a time-reversal-symmetry-breaking $s \pm is$ state, and eventually to a nodal $s_\pm$ state. The direction of the arrows inside the circle represent the relative phase $\theta$ of the superconducting order parameter and the length indicate the magnitude. **d,** Crystal structure of $Ba_{1-x}K_xFe_2As_2$. The cleaved As-terminated surface exhibits a (2 × 2) charge density wave modulation. **e,** Differential conductance map d$I$/d$V$ (r, $V$ = –10 meV) measured at 0.3 K [($V_s$, $I_s$) = − 10 mV, 200 pA; $V_{mod}$ =0.25mV]. Scale bar, 6 nm. **f,** Fourier transform of the differential conductance map shown in **e**. Red squares mark Bragg peaks of the (1 × 1) lattice, while blue squares denote the (2 × 2) CDW peaks. The zoomed-in view reveals a splitting of the CDW peaks—absent in the Bragg peaks—signifying the presence of CDW domain walls.

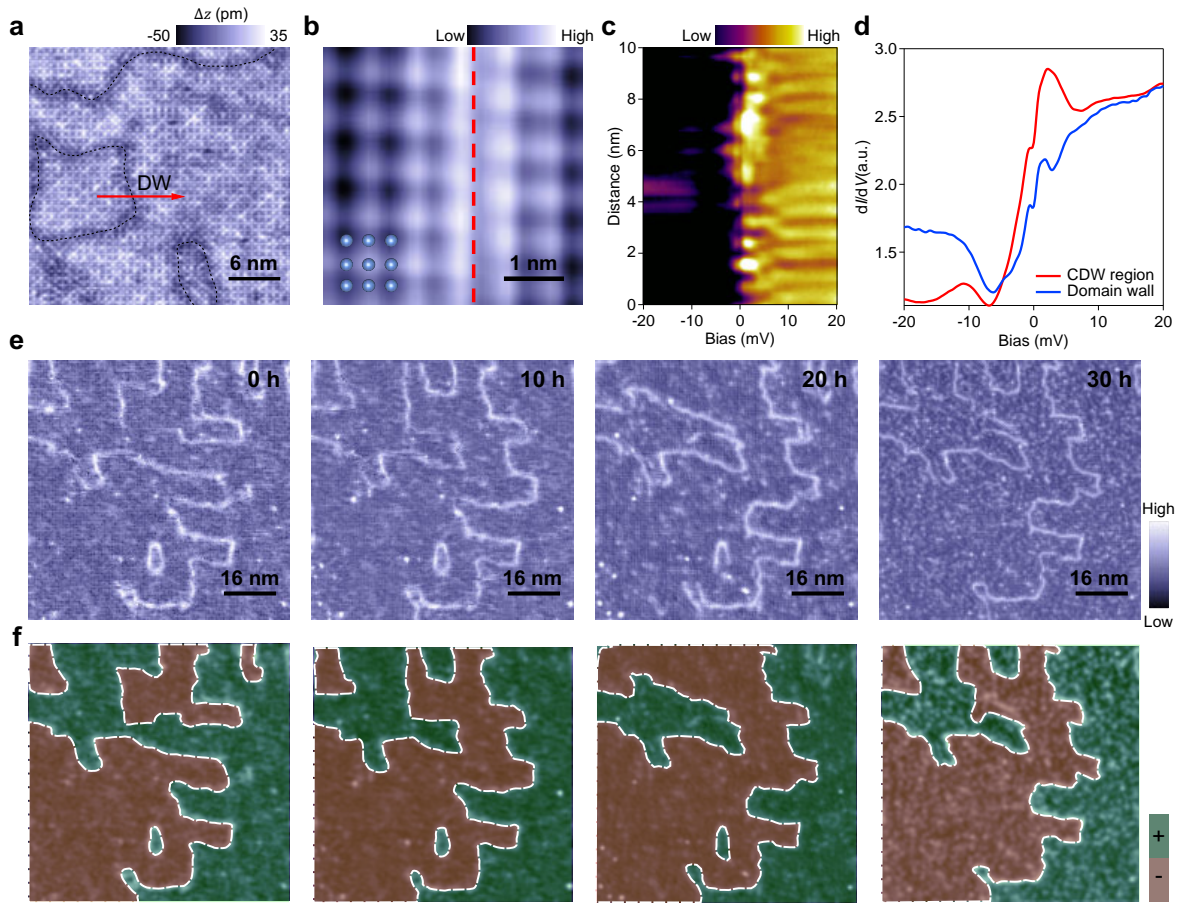

**Fig. 2 | Spectroscopic characterization and temporal evolution of domain walls. a,** Atomically-resolved topography of the As-terminated surface of $Ba_{0.23}K_{0.77}Fe_2As_2$. Black dashed lines mark the position of domain walls (DWs). [$(V_s, I_s)$ = (5 mV, 100 pA); Scale bar: 6 nm]. **b,** Enlarged differential conductance map dI/dV at -10 mV across a DW, overlaid with a ball model of surface As atoms, revealing a π-phase shift in the charge modulation. **c,** Spatially resolved dI/dV spectra acquired along the red line in **a** at 0.3 K [$(V_s, I_s)$ = − 10 mV, 200 pA; $V_{mod}$ =0.25mV]. **d,** Averaged dI/dV spectra taken from the CDW region (red curve) and the domain wall region (blue curve), respectively. **e,** Constant-current dI/dV maps recorded at different time on the As-terminated surface of $Ba_{0.23}K_{0.77}Fe_2As_2$ [$(V_s, I_s)$ = − 10 mV, 200 pA; $V_{mod}$ =0.75mV]. The DW distribution exhibits slow temporal drift, indicating dynamic behavior even at low temperature. **f,** Schematic illustration of two adjacent CDW domains, shown in red and green.

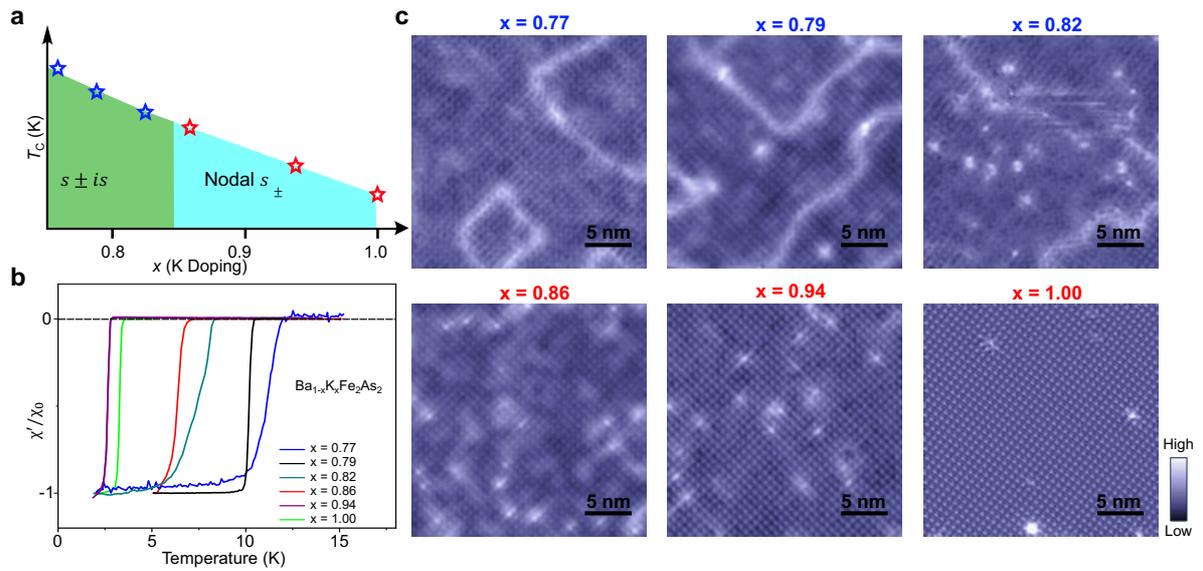

**Fig. 3 | Doping dependence of DWs. a, b,** AC susceptibility measurements were employed to determine the superconducting transition temperatures of $Ba_{1-x}K_xFe_2As_2$ single crystals with doping levels x = 0.77, 0.79, 0.82, 0.86, 0.94, and 1.00. Among them, the samples with x = 0.77, 0.79, and 0.82 (blue stars) lie within the regime of time-reversal symmetry-breaking superconducting pairing states, while the x = 0.86, 0.94, and 1.00 samples (red stars) exhibit nodal $s_\pm$ pairing state. **c,** Constant-current differential conductance maps $dI/dV_{cc}(\mathbf{r}, V = -10\text{mV})$ measured on the As-terminated surfaces of samples with different doping levels [$(V_s, I_s) = -10$ mV, 200 pA; $V_{mod}$ =0.75mV]. CDW is observed on the As-terminated surfaces of all doping levels, whereas DW features emerge exclusively in the TRSB doping regime.

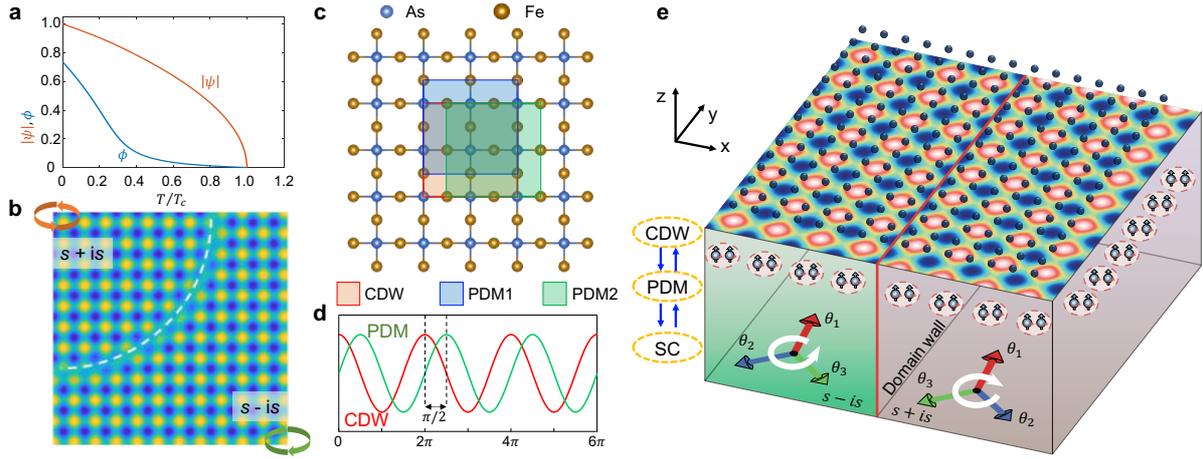

**Fig. 4 | Novel coupling between CDW and TRSB superconductivity. a,** Temperature dependence of uniform order parameters of superconductivity $|\psi|$ and charge order $\phi$ obtained by minimizing the GLW energy density with $T_c^{SC} > T_c^{CDW}$. For simplicity, an isotropic three-component superconductivity is presumed. **b,** GLW simulations reproduce a $\pi$-phase shift of the CDW across the domain wall separating the $s + is$ and $s - is$ superconducting domains. The white dashed line marks the superconducting domain boundary. **c,** $2 \times 2$ charge order on the top As atom layer and Cooper pair density modulations on the beneath atomic layer. CDW: $\rho_{CDW}(r) = \rho_0 + \phi[\cos(qx) + \cos(qy)]$, PDM1: $\psi_{PDW}(r) = \beta\psi_{Bulk}[\cos(qx) + \cos(q\{y + a/2\})]$, PDM2: $\psi_{PDW}(r) = \beta\psi_{Bulk}[\cos(q\{x + a/2\}) + \cos(qy)]$, with $q = \pi/a$. **d,** Schematic of the PDM: CDW inter-state discommensuration with a relative phase offset $\delta\varphi = \pi/2$. **e,** Schematic of the interplay between a surface charge-density wave (CDW) and bulk time-reversal-symmetry-breaking (TRSB) superconductivity. A hidden interfacial pair-density modulation (PDM) couples the bulk condensate to the surface CDW, thereby modulating the CDW phase and producing a $\pi$-phase shift across the domain wall that delineates macroscopic TRSB domains of opposite chirality. We refer to this state as a bipolar CDW.

# Methods

## Materials

The samples used in our STM measurements were selected according to their $T_c$ values. Phase purity and crystalline quality of plate-like $Ba_{1-x}K_xFe_2As_2$ single crystals were examined by X-ray diffraction and transmission electron microscopy. The $c$-axis lattice parameters were calculated from the X-ray diffraction data using the Nelson–Riley function. The doping level $x$ of K for the single crystals was determined using the relation between the $c$-axis lattice parameter and the K doping obtained in previous studies[49].

## Scanning tunnelling microscopy/spectroscopy

The STM/S experiments were performed using a commercial Unisoku USM1300 low-temperature STM machine that operates at a base temperature of 300 mK. Pt/Ir tips were used and conditioned by field emission on a gold target. To obtain clean surfaces for STM measurements, $Ba_{1-x}K_xFe_2As_2$ samples were cleaved in-situ at ~14 K in ultra-high vacuum (base pressure ~2 × $10^{-10}$ mbar), then immediately transferred to the STM stage (maintained at 4.2 K) for STM/S measurements. Conventional (constant height) differential conductance ($dI/dV$) spectra and maps $dI/dV$ (**r**, $V$) were recorded using a standard lock-in technique, with the frequency of bias modulation set at 973 Hz. Constant current differential conductance map $dI/dV_{cc}$ (**r**, $V$) slices were obtained through continuous acquisition of $dI/dV$ signals for every position of the scanned area in a STM scan at fixed scan set-points ($V_0$, $I_0$), during which the current feedback loop was kept closed. This method allows for acquisition of differential conductance map slices in a relatively short time, and reveal the same features as $dI/dV$(**r**, $V$) map slices obtained by conventional means[18].

## Ginzburg-Landau-Wilson energy

We first construct a phenomenological model to simulate and reproduce the experimentally observed phenomena. A detailed Ginzburg-Landau-Wilson energy density for the present system can be given as:

$$H = \sum_{j=1,2,3}\left[a_j|\psi_j|^2 + \frac{b_j}{2}|\psi_j|^4 + \frac{1}{2m_j}\left|\frac{\hbar}{i}\nabla\psi_j\right|^2\right] - \sum_{j,k=1,2,3;j<k}\gamma_{jk}(\psi_j^*\psi_k + \text{c.c.})$$
$$+a'\phi^2 + \frac{b'}{2}\phi^4 + \frac{\hbar^2}{2m'}(\nabla\phi)^2 + \alpha\phi^2\sum_{j=1,2,3}|\psi_j|^2 + i\beta\phi[\psi_1^*\psi_2 - \psi_1\psi_2^* + (1,2,3 \text{ cyclic})], \quad 3)$$

where $\prod_{\langle jk\rangle}\gamma_{jk} < 0$ establishes TRSB superconductivity[28]. $\psi_j = |\psi_j|e^{i\theta_j}$ are complex order parameters representing the three superconducting condensates labelled by the indices $j$=1, 2, 3.

For simplicity, an isotropic three-component superconductivity is presumed: $a_j = a_0(T/T_c^{SC} - 1)$, $b_j = b$, $m_j = m$, $\gamma_{jk} = \gamma$ with $a_0, b, m > 0$ and $\gamma < 0$, whereas $a' = a_0'(T/T_c^{CDW} - 1)$ with $a_0', b', m' > 0$, and $T_c^{SC} > T_c^{CDW}$, $\alpha$ and $\beta$ real numbers. Typical temperature dependence of order parameters is displayed in Fig. 4a. Charge order exhibits a temperature dependence that contrasts sharply with that of superconductivity. The superconducting order parameter onsets with an amplitude scaling as $\sqrt{1 - T/T_c}$ when temperature decreases to a transition point $T_c$ given by $a_0(T_c/T_c^{SC} - 1) + \gamma = 0$ [28]. By contrast, the charge-order intensity is rapidly suppressed upon heating and, in the vicinity of the same critical temperature $T_c$, grows linearly from zero, reflecting the linear coupling between the two orders (see Eq. (3)). Over an extended temperature interval well below $T_c$, the charge-order amplitude remains too weak to be resolved experimentally, consistent with previous reports for the same material[18]. Recently a coupling $iM_z(\psi_1^*\psi_2 - \psi_1\psi_2^*)$ was discussed between a multi-component superconductivity and an antiferromagnetic order in nonsymmorphic structures[50]. While it is similar in form to Eq. (3), be noticed that their antiferromagnetic order $M_z$ breaks TRS, while the present charge order $\phi$ does not.

In order to demonstrate directly the locking between the charge order and TRSB superconductivity, we perform numerical calculations for the situation where two TRSB superconductivity states coexist in the system and are separated by an interface by imposing appropriate boundary conditions on superconductivity, and check the stable wavefunction of the whole system which minimizes the (free-)energy density in Eq. (3). When two phase variables of superconductivity order parameters cross each other (see Extended Data Fig. 8), which switches the $s + is$ superconducting state into the $s - is$ one, a sign change in the CDW order parameter occurs. As displayed in Figs. 4b, accompanying the reversal between the two TRSB superconductivity states, the 2×2 charge orders on the two sides of the interface slide by a half-wavelength. This fully agrees with the experimental observations unveiled above, indicating compellingly that the novel coupling between the charge order and the TRSB superconductivity given in Eq. (3) takes place in the highly doped $Ba_{1-x}K_xAsFe$ superconductor.

**Dimensionless quantities, numerical parameters and numerical techniques**

For numerical calculations based on the GLW energy density in Eq. (3), dimensionless quantities are used as follows: $\psi_j = \tilde{\psi}_j \times \sqrt{a_0/b}$, $\phi = \tilde{\phi} \times (a_0/b)$, where $\phi$ is taken as number density of charge in units of $2e$, $x, y = \tilde{x} \times \xi, \tilde{y} \times \xi$, where $\xi = \hbar/\sqrt{2ma_0} = 2a_{As}$ is set as twice of the lattice constant, $\gamma = \tilde{\gamma} \times a_0$, $a_0' = \tilde{a}_0' \times b$, $b' = \tilde{b}' \times (b^3/a_0^2)$, $\alpha = \tilde{\alpha} \times (b^2/a_0)$, $\beta = \tilde{\beta} \times b$, $\eta = \tilde{\eta} \times (b\xi^2)$, $\tilde{a} = T/T_c^{SC} - 1$, and $\tilde{a}' = \tilde{a}_0' \times (T/T_c^{CDW} - 1)$. For the numerical results displayed in Fig. 4a where tildes are omitted for simplicity, parameters are taken as $\tilde{\gamma} = -0.1$, $\tilde{a}_0' = 0.8$, $\tilde{b}' = 1$, $\tilde{\alpha} = 0.1$, $\tilde{\beta} = 0.03$, $\tilde{\eta} = 1$, and $T_c^{CDW} = T_c^{SC}/3$. Numerical simulations are conducted via the finite difference method. Results are obtained through an iterative relaxation technique based on Eq. (3). To obtain coexisting domains of opposite superconductivity chirality, a suitable initial configuration of the order parameters is required.

**Extraction of amplitude and phase maps**

We use a two-dimensional lock-in technique (demodulate-then-low-pass) to extract the amplitude and phase of the charge density map. We first compute the centered FFT of $g(\mathbf{r}) = dI/dV (\mathbf{r}, V = -10$ meV$)$ and locate each bipolar CDW central peak $\mathbf{q}_j^{CDW}$, $(j = 1, 2)$ by sub-pixel 2D-Gaussian fitting of the local maximum. The real-space image is then demodulated by the refined carrier, $g(\mathbf{r}) \to g(\mathbf{r}) e^{-i\mathbf{q}_j^{CDW} \cdot \mathbf{r}}$, which shifts the $+\mathbf{q}_j$ sideband to $\mathbf{q} = 0$. To suppress spectral leakage from finite boundaries we apply a mild apodization in real space (Hann window), and then remove the spatial mean to avoid DC contamination after demodulation. In Fourier space we apply a centered Gaussian low-pass $G_\sigma(\mathbf{q}) = \exp[-|\mathbf{q}|^2/(2\sigma^2)]$. The width $\sigma$ is chosen large enough to fully encompass the CDW peak around $\pm\mathbf{q}_j^{CDW}$, avoiding artificial truncation of the physical signal, yet small compared with $|\mathbf{q}_j^{CDW}|$ so that residual components near $\pm 2\mathbf{q}_j^{CDW}$ are strongly attenuated. Considering the length scale (7-8 nm$^{-1}$) of the CDW modulations, we use a reasonable Gaussian width here ($\sigma = 2.5$ nm$^{-1}$). Under this Gaussian low-pass filter, oscillation components with $|\mathbf{q}| \gtrsim \sigma$ are exponentially suppressed, leaving only the slowly varying envelope encoded in the local phase $\varphi_j(\mathbf{r})$. The inverse FFT yields the complex lock-in field

$$\rho^0_{q_j^{CDW}}(\mathbf{r})e^{i\varphi_j(r)} = \mathcal{F}^{-1}\left\{\mathcal{F}\left[g(\mathbf{r})e^{-i\mathbf{q}_j^{CDW}\cdot\mathbf{r}}\right]G_\sigma(\mathbf{q})\right\}, \qquad 4)$$

from which we extract the phase and amplitude maps as the argument and magnitude of the field.

For visualization, we wrap the raw phase into the principal interval $(-\pi, \pi)$ by applying $\varphi_j(\mathbf{r}) \to \mathrm{mod}[\varphi_j(\mathbf{r}), 2\pi] - \pi$, which removes artificial $2\pi$ jumps and makes the $2\pi$-winding topological defects directly visible in Extended Fig. 3c and 3g. To quantify spatial phase variations, however, we compute the phase gradient $\nabla\varphi_j(\mathbf{r})$ on the unwrapped phase maps. The phase gradient map shown in Extended Fig. 3d and 3h is the magnitude of this vector field $|\nabla\varphi_j(\mathbf{r})|$. Domain walls correspond to lines across which the CDW phase changes abruptly by $\pi$, so that the local phase varies much more rapidly there than within a single domain. As a result, $|\nabla\varphi_j(\mathbf{r})|$ develops sharp, filamentary maxima along these phase-slip lines, and the bright trajectories in the phase-gradient maps provide a direct, quantitative visualization of the domain walls.

## Acknowledgements

H.D. acknowledges support from the New Cornerstone Science Foundation (No. 23H010801236), Innovation Program for Quantum Science and Technology (No. 2021ZD0302700), National Natural Science Foundation of China (No.12488101). X.H. acknowledges support from the Shanghai Science and Technology Innovation Action Plan (No. 24LZ1400800). V.G. acknowledges support from Natural Science Foundation of China (No.12374139 and No.12350610235). B.L. acknowledges support from the National Natural Science Foundation of China (No. 12374063), the Ministry of Science and Technology of China (2023YFA1407400), the Shanghai Natural Science Fund for Original Exploration Program (23ZR1479900), and the Cultivation Project of Shanghai Research Center for Quantum Sciences (Grant No. LZPY2024). Q.H. acknowledges support from China Postdoctoral Science Foundation (No. GZB20230421).

## Author contributions

Q.H. and H.D. conceived the experiment. Q.H. performed STM measurements with the assistance of Q.W. and X.L.. Q.H. performed the data analysis with assistance of Q.W.. A portion of STM data were collected by Q.H. with the assistance of Y.Z.. L.Z., K.T. and X.H. composed the theory and performed numerical simulations. Y.L. and V.G. prepared and characterized (or selected) samples for the research. Q.H., X.H. and H.D. wrote the paper with contributions from all authors. H.D. supervised the project.

## Data and materials availability:

The data that support the findings of this study are available upon reasonable request.

## Competing interests:

Authors declare that they have no competing interests.

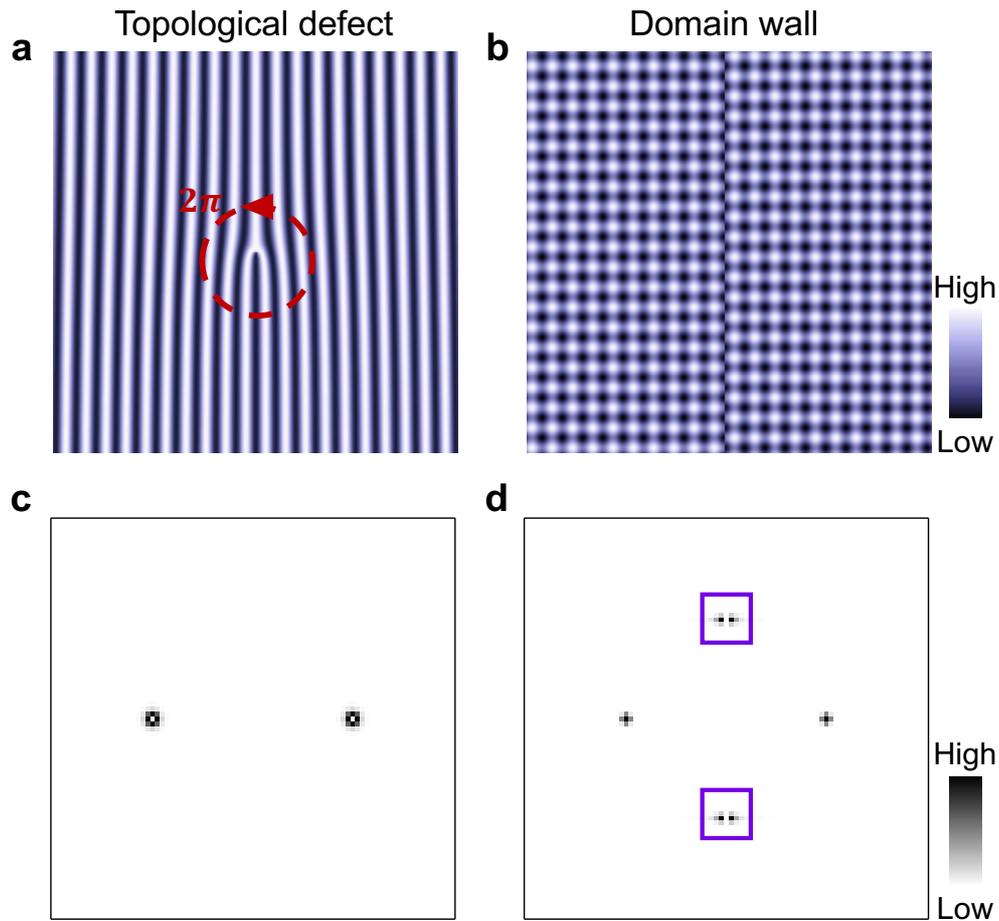

**Extended Data Fig. 1 | Simulations of topological defects and domain wall in CDW. a,** Schematic of a dislocation/topological defect in an ordered CDW texture, around which the phase winds by $2\pi$. **b,** Schematic of CDW domain wall (DW). Crossing the DW, the CDW intensity inverts, switching from a maximum to a minimum. **c, d,** Fourier transform corresponding to **a** and **b,** respectively. In reciprocal space, the CDW Bragg peaks remain single for the isolated dislocation (**c**), whereas inclusion of a DW within the field of view produces a characteristic splitting of each CDW Bragg peak into two symmetrically displaced components (indicated by the purple boxes) (**d**).

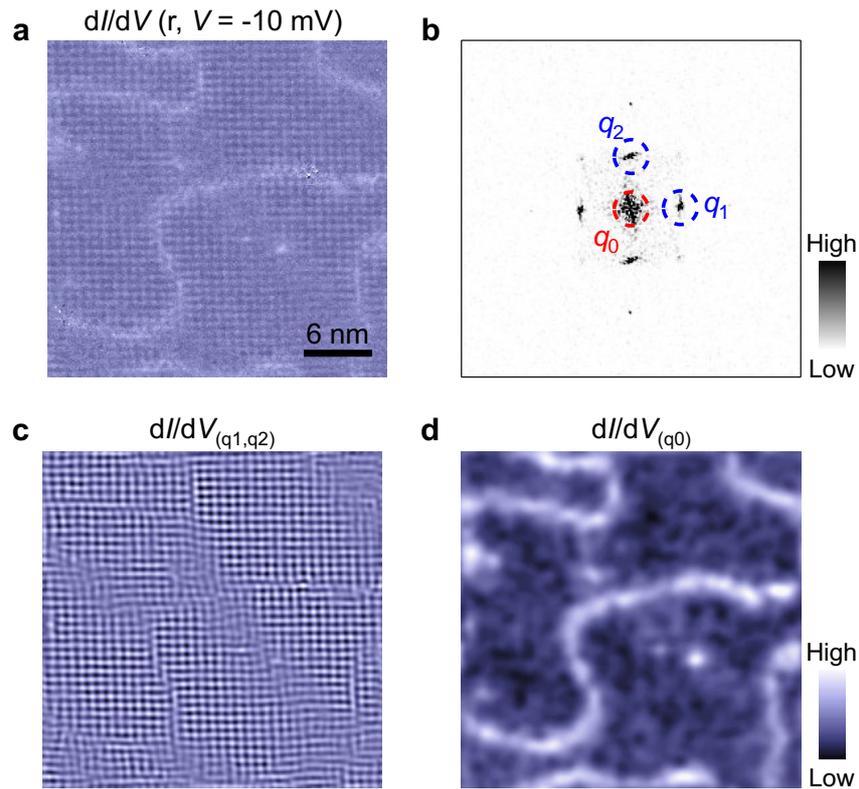

**Extended Data Fig. 2 | Fourier transform (FT)-filtered image. a,** Differential conductance map d$I$/d$V$ (r, $V$ = –10 meV) measured at 0.3 K [($V_s$, $I_s$) = − 10 mV, 200 pA; $V_{mod}$ =0.25mV]. **b,** Fourier transform of the differential conductance map shown in a. c, d, FT-filtered images at ($q_1$, $q_2$) (**c**) and $q_0$ (**d**) of **b**. The FT-filtered image at $q_0$ (**d**) clearly reveals the spatial distribution of domain walls.

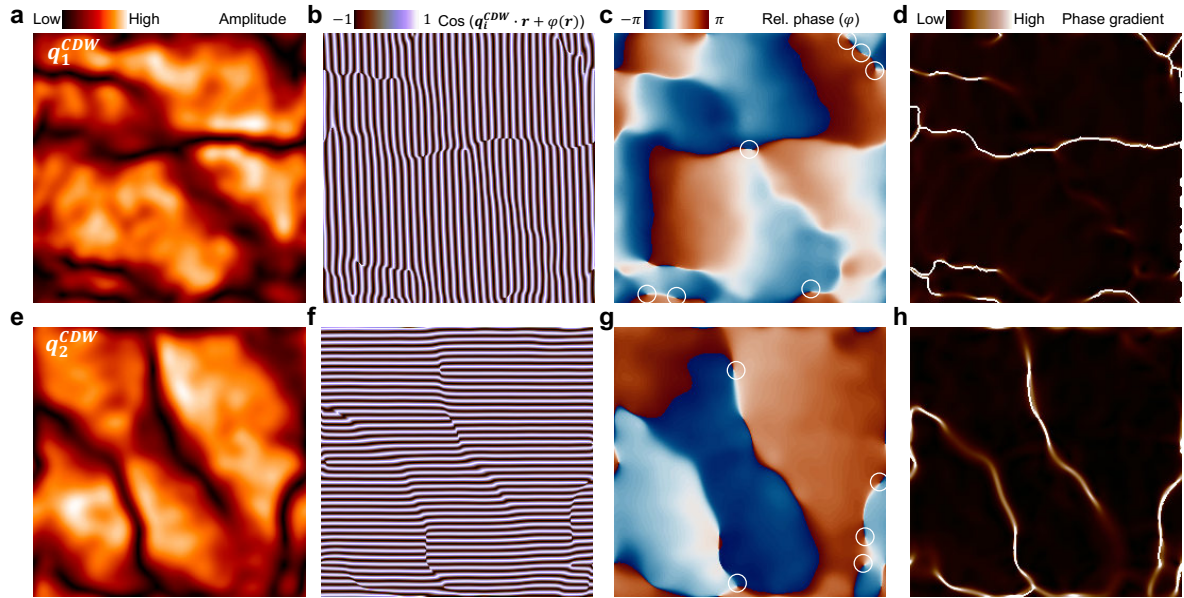

**Extended Data Fig. 3 | Extraction of amplitude and phase maps. a-d,** For the CDW component at $q_1^{CDW}$: amplitude **(a)**, $\cos(q_1^{CDW} \cdot r + \varphi(r))$ **(b),** relative phase $\varphi$ **(c)** and phase gradient **(d)** maps. **e-h,** Corresponding maps for the orthogonal component at $q_2^{CDW}$: amplitude **(e)**, $\cos(q_2^{CDW} \cdot r + \varphi(r))$ **(f),** relative phase $\varphi$ **(g)** and phase gradient **(h)** maps.

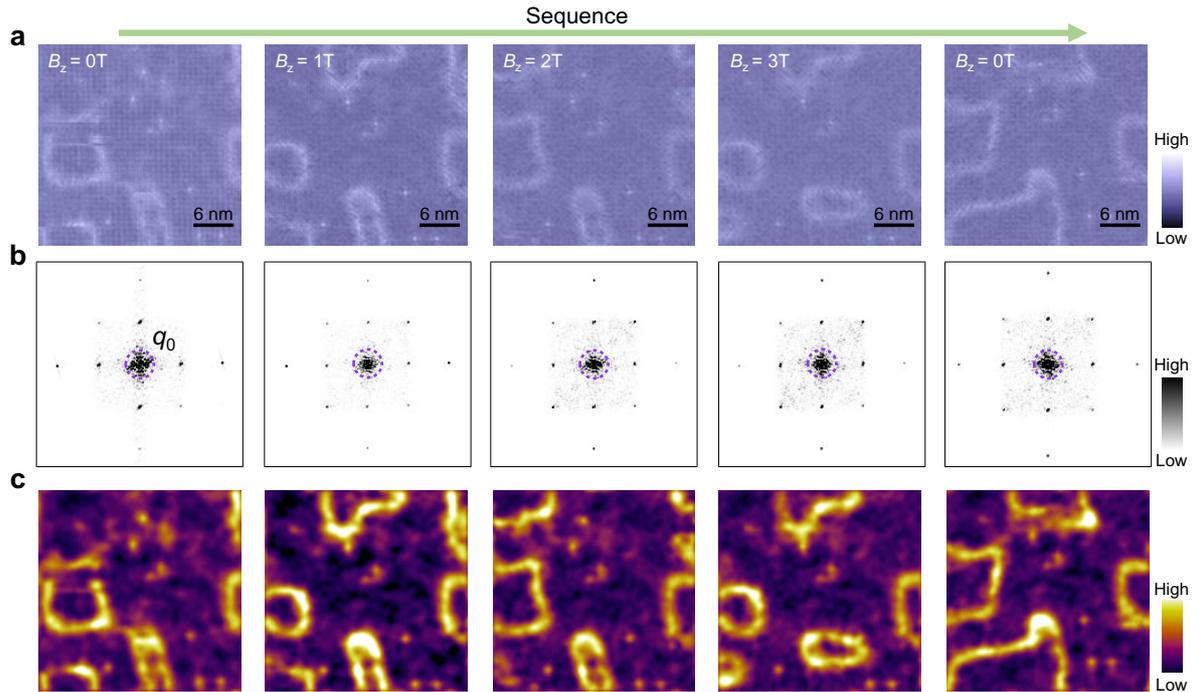

**Extended Data Fig. 4 | Magnetic field dependent measurements. a,** Constant-current differential-conductance maps $dI/dV_{cc}$ (r, V) acquired over the same field of view under different out-of-plane magnetic fields. **b,** Fourier transforms of the maps in **a**. **c,** FT-filtered reconstructions at $q_0$, as marked by purple circles in **b**. The spatial distribution of the DWs is highly sensitive to magnetic field and exhibits pronounced magnetic-history dependence: after cycling the field back to zero, the DW configuration differs markedly from the initial zero-field state, indicating a strong dependence on the sequence of field application.

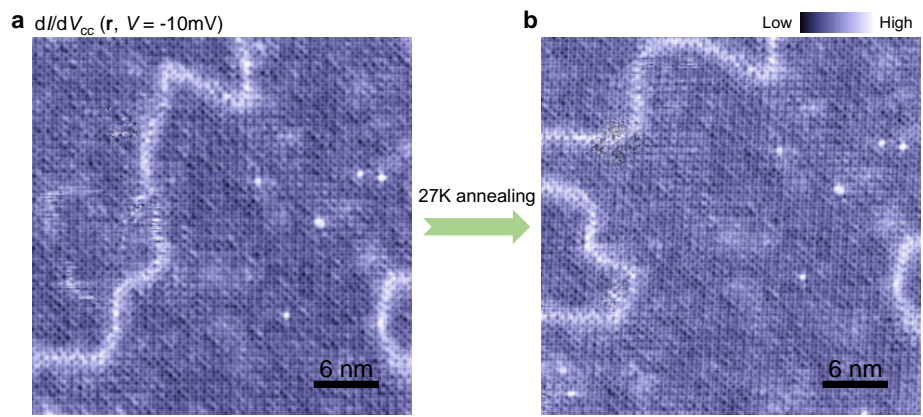

**Extended Data Fig. 5 | Thermal annealing. a,** Constant-current differential-conductance maps d$I$/d$V_{cc}$ (r, V) acquired at $T$=1.8 K. **b,** The same field of view re-measured at $T$=1.8 K after thermal annealing at 27 K (above the bulk superconducting transition temperature). The domain-wall configuration is completely reconfigured relative to **a**, exhibiting a markedly different spatial distribution and indicating strong thermal-history dependence.

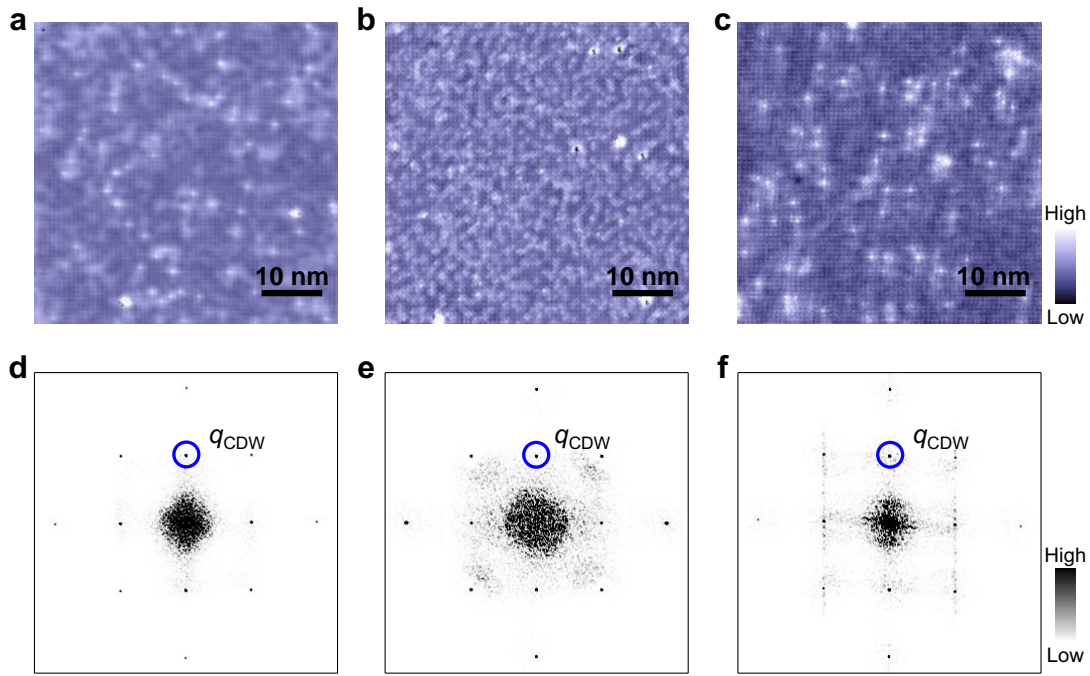

**Extended Data Fig. 6 | Different regions of Ba$_{0.14}$K$_{0.86}$Fe$_2$As$_2$. a-c,** Constant-current differential-conductance maps d$I$/d$V_{cc}$ (r, V) acquired at three distinct regions of Ba$_{0.14}$K$_{0.86}$Fe$_2$As$_2$. **d-f,** Fourier transforms of the maps in **a-c**, respectively. Across multiple As-terminated surfaces at this doping, we observe clear CDW contrast (blue circles) but no domain walls.

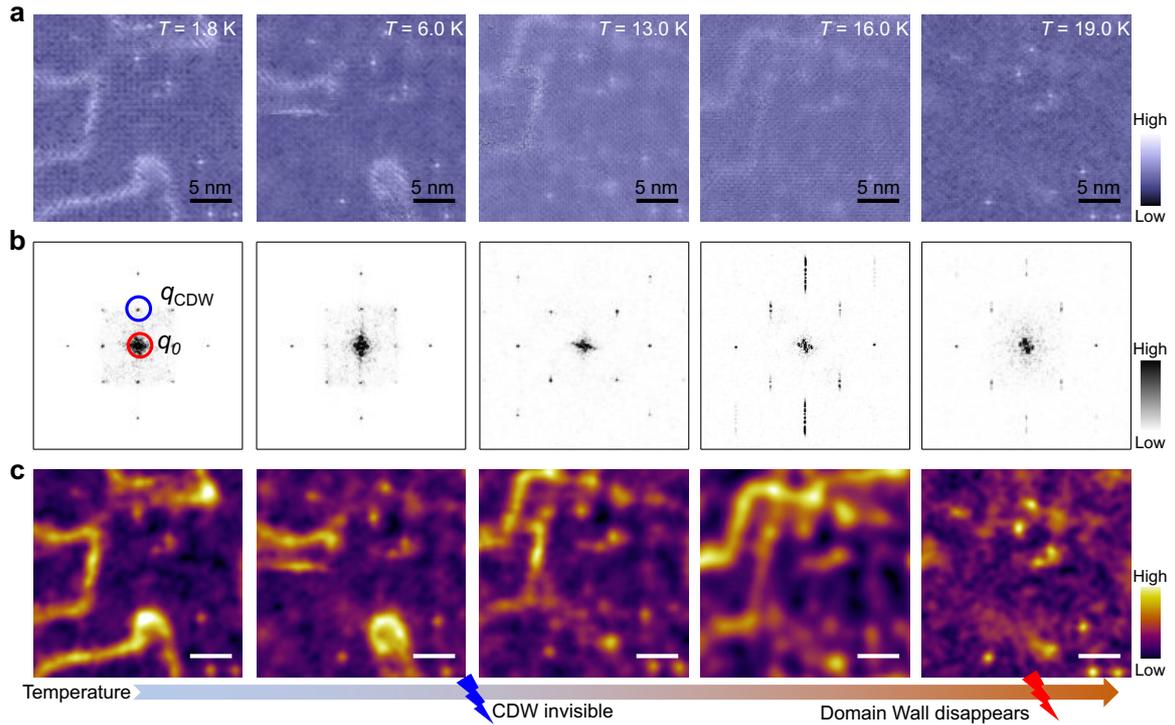

**Extended Data Fig. 7 | Temperature dependence of CDW and DWs in $Ba_{0.23}K_{0.77}Fe_2As_2$.** **a,** Constant-current differential-conductance maps $dI/dV_{cc}$ (r, V) acquired over the same field of view at different temperatures. **b,** Fourier transforms of the maps in **a**. **c,** FT-filtered reconstructions at $q_0$ (red circles in **b**). As temperature increases, the CDW contrast is rapidly suppressed and becomes indiscernible by 13 K, whereas the domain walls remain robust and persist up to 19 K.

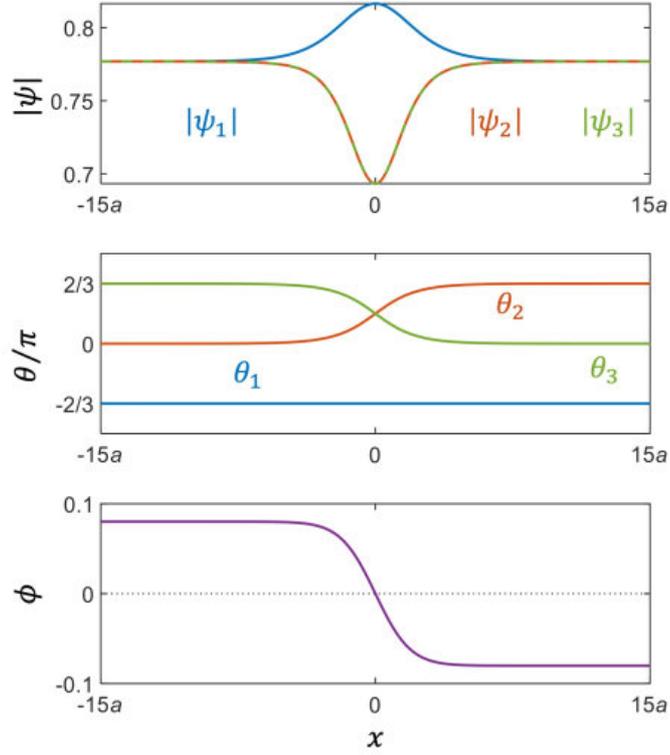

**Extended Data Fig. 8 | Structure of domain wall where $s + is$ and $s - is$ superconducting states are separated and CDW order parameter changes sign simultaneously. top,** Profiles of absolute values of the three superconductivity order parameters obtained by GLW simulations at a typical low temperature. Two of them are suppressed and the remaining one is enhanced at the domain wall, corresponding to a unique superconductivity collective mode. **middle,** Profiles of phases of the three superconductivity order parameters. Two of them cross each other at the domain wall, which switches the $s + is$ superconducting state into the $s - is$ one. **bottom,** Profile of the CDW order parameter. The CDW order parameter changes sign at the domain wall, indicating a half-wavelength sliding in the 2×2 CDW order. For the numerical results displayed here, dimensionless quantities are used (for simplicity tilt mark is not shown explicitly), and parameters are taken same as those in Fig. 4a in the main text, namely $\xi = 2a_{As}$, $\tilde{\gamma} = -0.1$, $\tilde{a}'_0 = 0.8$, $\tilde{b}' = 1$, $\tilde{\alpha} = 0.1$, $\tilde{\beta} = 0.03$, $\tilde{\eta} = 1$, and $T_c^{CDW} = T_c^{SC}/3$, and a is set as the lattice constant $a_{As}$.

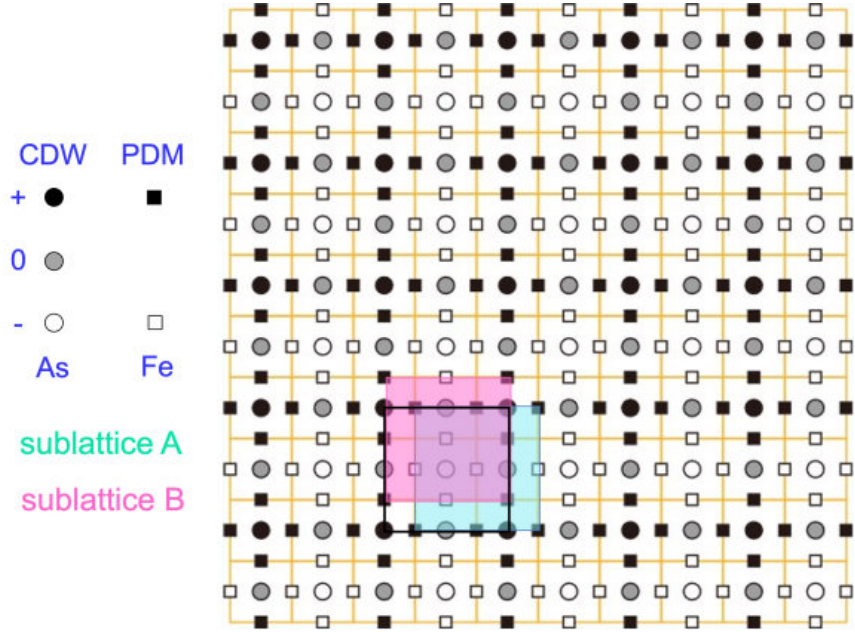

**Extended Data Fig. 9 | Schematic illustration of the 2×2 CDW order at the top As layer and the hidden PDM on the beneath Fe layer.** The CDW order (circles) is given by $\rho_{\text{CDW}}(r) = \cos(qx) + \cos(qy)$, whereas the two PDM orders (squares) are given by $\psi_{\text{PDM}}(r) = \cos(q\{x + a/2\}) + \cos(qy)$ for the A sublattice, and $\psi_{\text{PDM}}(r) = \cos(qx) + \cos(q\{y + a/2\})$ for the B sublattice, respectively, with $q = \pi/a$. Black/gray/white circles are for +/0/- value of CDW order, black/white squares are for +/- value of PDM order.

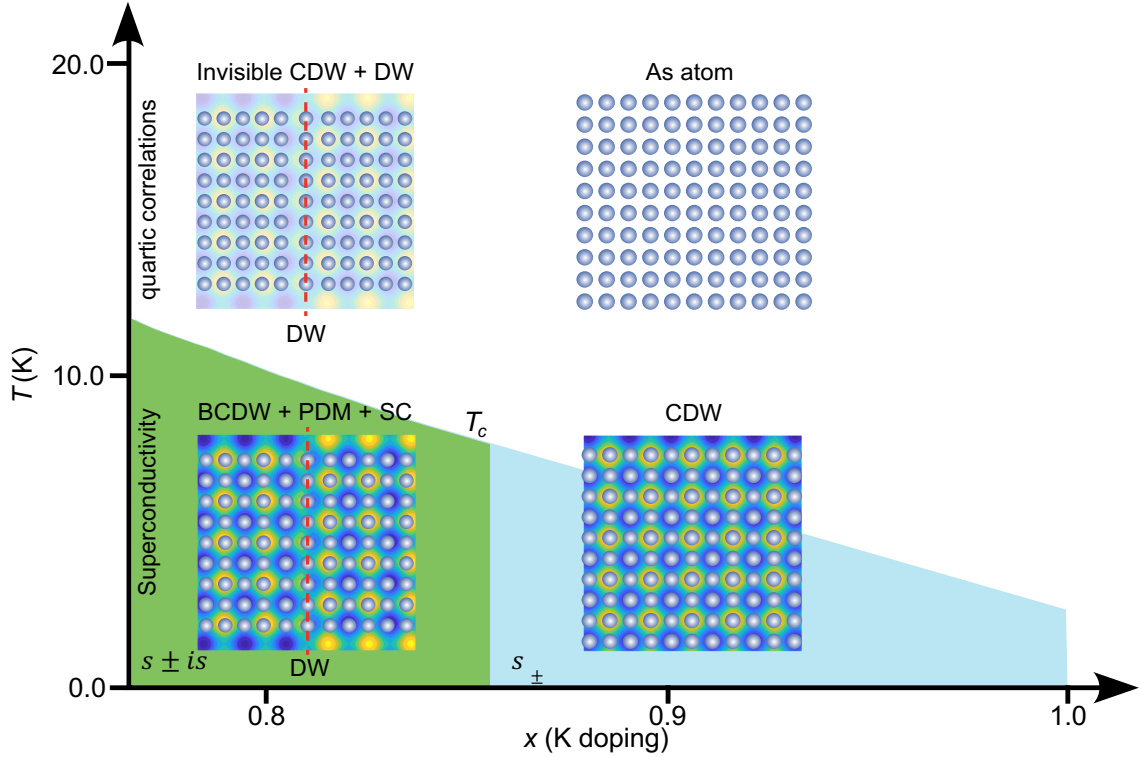

**Extended Data Fig. 10 | Evolution of the charge order and superconductivity in heavily hole doped $Ba_{1-x}K_xFe_2As_2$.** In the heavily hole-doped regime, the As-terminated surface develops a $2 \times 2$ checkerboard-like CDW, which suppresses superconductivity on the surface while leaving the bulk superconducting state intact. Notably, within the "magic doping" region ($0.70 < x < 0.85$), the bulk superconducting pairing state breaks $U(1) \times Z_2$ symmetries. In this doping range, the TRSB superconductivity imparts "chirality" to the CDW, yielding commensurate domains separated by domain walls (DW) across which the CDW phase slips by $\pi$—forming what we term a bipolar charge density wave (BCDW). A hidden interfacial pair-density modulation (PDM) is required to mediate a linear coupling between the surface CDW and the TRSB superconductivity, enabling the formation of the BCDW and its DWs. Consistent with this linear coupling, the CDW contrast is rapidly suppressed with increasing temperature whereas the DW signal persists to higher temperatures (up to ~ 20 K), indicative of quartic correlations that survive into the fluctuating regime above $T_c$.